\documentclass[prl,twocolumn,amsmath,amssymb,showpacs]{revtex4}
\usepackage{graphicx}% Include figure files
\usepackage{dcolumn}% Align table columns on decimal point
\usepackage{bm}% bold math
\usepackage{mathrsfs}
\usepackage{appendix}
\usepackage{bbm}
\usepackage{color}

\def \uudd {\uparrow \uparrow \downarrow \downarrow}
\def \vQ {{\bf Q}}
\def \vP {{\bf P}}
\def \vS {{\bf S}}
\def \vR {{\bf R}}
\def \vG {{\bf G}}

\def \vq {{\bf q}}
\def \vx {{\bf x}}
\def \vy {{\bf y}}

\def \sgn {{\rm sgn}}

\begin{document}

\title{Non-Collinear Magnetic Phases of a Triangular-Lattice Antiferromagnet and Doped CuFeO$_2$}
\author{Randy S. Fishman and Satoshi Okamoto}
\affiliation{Materials Science and Technology Division, Oak Ridge National Laboratory, \mbox{Oak Ridge, TN 37831}}

\begin{abstract}

We obtain the non-collinear ground states of a triangular-lattice antiferromagnet with exchange interactions
up to third nearest neighbors as a function of the single-ion anisotropy $D$.  At a critical value of $D$, the collinear $\uudd $ phase 
transforms into a complex non-collinear phase with odd-order harmonics of the fundamental ordering wavevector
$\vQ $.  The observed elastic peaks at $2\pi \vx -\vQ $ in both Al- and Ga- doped CuFeO$_2$ are explained by
a ``scalene" distortion of the triangular lattice produced by the repulsion of neighboring oxygen atoms.

\end{abstract}


\maketitle

The non-collinear and multiferroic ground states of frustrated magnetic systems continue to attract intense interest.
Due to the strong coupling between the electric polarization and the non-collinear spin states, improper ferroelectric materials
hold great technological promise \cite{imp}.  However, more than one physical mechanism may be responsible for their ferroelectric
behavior.   An electric polarization $\vP $ perpendicular to both the spin rotation axis $\vS_i \times \vS_j$ and the
wavevector $\vQ $ is predicted for ferroelectrics with easy-plane anisotropy and spiral spin states like $R$MnO$_3$ 
($R$ = Tb or Y) \cite{spiral}.  But a simple spiral state is not possible for ferroelectrics based on materials
with easy-axis anisotropy, like CuFeO$_2$ \cite{Nak07, Nak08, Terada08} and MnWO$_4$ \cite{mnw}.  
For Al- or Ga- doped CuFeO$_2$, a modulation of the 
metal-ligand hybridization with the spin-orbit coupling \cite{Jia06, Arima07} may produce the observed electric 
polarization $\vP $ \cite{Nak07, Nak08, Terada08} parallel to both the spin rotation axis and 
the wavevector.  

In order to clarify the nature of the ferroelectric coupling, it is essential to understand how the non-collinear ground 
state of an easy-axis ferroelectric evolves with doping.  In this paper, we show that the 
non-collinear ground state of CuFeO$_2$ contains significant odd-order harmonics of the fundamental ordering wavevector 
$\vQ \approx 0.86 \pi \vx $ \cite{thr}.  The observed elastic peaks at both $\vQ $ and $2\pi \vx -\vQ \approx 1.14 \pi \vx $ 
\cite{Nak07, Terada08} are explained by a distortion of the triangular lattice associated with the repulsion of 
neighboring oxygen atoms.

Due to geometric frustration, simple antiferromagnetic (AF) order is not
possible on a two-dimensional triangular lattice with AF interactions $J_1< 0 $ between 
neighboring sites.  When the easy-axis anisotropy $D$ along the $z$ axis is sufficiently large, however,
the anisotropy energy $-D\sum_i {S_{iz}}^2$ favors one of several collinear states.  For classical spins, 
Takagi and Mekata \cite{Takagi95} demonstrated 
that the $\uudd$ state sketched in Fig.1 is stable over the range of $J_2/\vert J_1\vert $ and $J_3/\vert J_1\vert $ 
plotted in the inset to Fig.2(a), where $J_2$ and $J_3$ are the second- and third-neighbor interactions 
indicated in Fig.1 and longer-ranged interactions are neglected.  The $\uudd $ phase with wavevector 
$\vQ_0 =\pi \vx $ appears in pure CuFeO$_2$ \cite{Mitsuda91, Mekata93} for magnetic fields below about 7 T.  

With increasing Al concentration, the spin-waves (SWs) of CuFe$_{1-x}$Al$_x$O$_2$ soften on either side of $\vQ_0$
at wavevectors $\vQ_{\pm }  \approx \vQ_0 \pm 0.14 \pi \vx $ \cite{Terada04, Ye07}.   A similar SW softening occurs on a triangular 
lattice AF when $D$ is lowered while the exchange constants are fixed \cite{Fish08}.  
For an Al concentration $x$ greater than $x_c \approx 0.016$ or an anisotropy $D$ lower than 
$D_c \approx 0.3\vert J_1\vert $, 
the $\uudd $ phase becomes unstable and a non-collinear phase appears \cite{Kan07, Seki07} with the
dominant wavevector $\vQ \equiv \vQ_-\approx 0.86 \pi \vx$.  

\begin{figure}
\includegraphics *[scale=0.52]{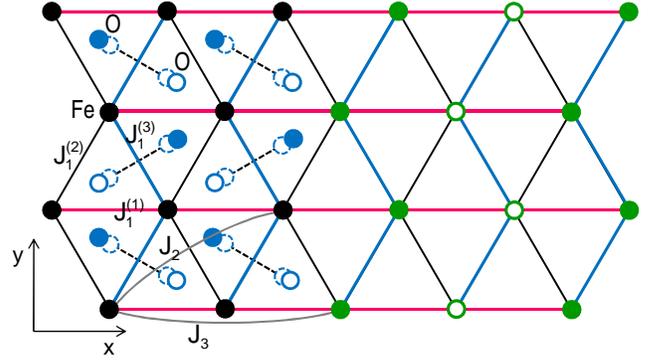}
\caption{
(Color online) The exchange constants $J_1^{(i)}$, $J_2$, and $J_3$, and the oxygen displacements
responsible for the scalene distortion of the lattice.  The thick
bonds $J_1^{(3)}=J_1+K_1$ form a zig-zag pattern.  Also shown is the $\uudd $ state with $\uparrow $ (solid)
and $\downarrow $ (open) spins.  For $K_1 > 0$, $J_1^{(3)}$ couples the same spin along each zig-zag.
}
\end{figure}

Because the AF interactions between adjacent
hexagonal layers of CuFeO$_2$ are not frustrated, the essential
physics of CuFeO$_2$ is captured by a two-dimensional triangular-lattice AF
with energy 
\begin{equation}
\label{Ham}
E=-\frac{1}{2}\sum_{i\ne j}J_{ij}\vS_i \cdot \vS_j -D\sum_i {S_{iz}}^2.
\end{equation}
A classical approximation for the $S=5/2$ spins of the Fe$^{3+}$ ions incurs only small errors, so
each spin $\vS_i=\vS (\vR_i )$ is treated classically with $\vert \vS (\vR_i) \vert =1$.  We include exchange couplings 
$J_{ij}$ up to third nearest neighbors.  

Monte-Carlo simulations were recently used to study the complex non-collinear (CNC) phase of Eq.(\ref{Ham}) \cite{Har09}.
Those simulations indicated that the CNC phase interceded between the $\uudd $ phase at high $D $ 
and a spiral phase at small $D$.  Several ``trial" spin functions were constructed
to minimize the energy of Eq.(\ref{Ham}), including functions with all three spin components.  Those trial
functions were motivated by the Fourier peaks in the Monte-Carlo solution \cite{Har09}
at wavevectors $\vQ = 0.87 \pi \vx $, $1.13 \pi \vx = 2 \pi \vx - \vQ$, and $1.38 \pi \vx \approx -3\vQ +\vG $ where
$\vG =4\pi \vx $ is a reciprocal lattice vector.

The trial function with the lowest energy contains odd-order harmonics 
in an expansion of the spin:
\begin{eqnarray}
\label{sp1}
S_z(\vR )&=&A\Bigl\{ \cos(Qx) + \sum_{l=1} C_{2l+1} \cos \bigl(Q(2l+1)x \bigr) \nonumber \\ & 
+& \sum_{l=0} B_{2l+1} \cos \bigl(Q' (2l+1)x + \phi \bigr)\Bigr\} ,
\end{eqnarray}
\begin{equation}
\label{sp2}
S_y(\vR ) = \sqrt{1-S_z(\vR )^2}\, \sgn \bigl(\sin (Qx) \bigr),
\end{equation}
where $\vQ' =2\pi \vx  -\vQ $.   Notice that $\vS (\vR )= \vS (x)$ depends only on $x$.
The anharmonic coefficients $C_{2l+1 > 1}$ reflect the deviation from 
a pure cycloid with $\vS (x ) = \bigl( 0, \sin (Qx), \cos (Qx)\bigr)$;  
the coefficients $B_{2l+1}$ are produced by a lattice distortion
with period 1, as discussed further below.  The amplitude $A$ is fixed by the constraint that 
max$\vert S_z(x )\vert =1$ and the lattice constant is set to 1.  Keep in mind that $\vQ $ and
$\vQ'$ are distinct wavevectors not related by a reciprocal lattice vector.  

Like the observed
multiferroic phase \cite{Nak07}, the CNC phase of Eqs.(\ref{sp1}) and (\ref{sp2}) is coplanar with 
the spin rotation axis $\vS (x ) \times \vS (x+1/2)$ 
parallel to $\vQ $ along the $x$ axis.  A uniform rotation of 
$\vS (x )$ about the $z$ axis would not cost any anisotropy energy but would cost 
magneto-elastic energy due to the distortions discussed below.

As shown in Ref.\cite{Har09}, the dominant wavevector of the CNC phase coincides with the wavevector
of the dominant SW instability of the $\uudd $ phase.  Depending on whether the exchange parameters
$\{J_2/\vert J_1\vert ,J_3/\vert J_1\vert \}$ fall within regions 4I or 4II plotted in the inset to Fig.2(a), the dominant SW
instabilities occur at a variable wavevector $\vQ$ (region 4I) or at $(4\pi /3 )\vx $ (region 4II) \cite{Swan09}.   
The exchange parameters used in Fig.2(a) ($J_2/\vert J_1 \vert =- 0.20$, $J_3/\vert J_1\vert = -0.26$) fall
within region 4II;  the exchange parameters used in Fig.2(b) ($J_2/\vert J_1 \vert =- 0.44$, $J_3/\vert J_1\vert = -0.57$)
fall within region 4I.  The latter are believed to correspond approximately to the ratio of
exchange parameters in pure CuFeO$_2$ \cite{Ye07}.
With those parameters, the dominant SW instability of the $\uudd $ phase and the dominant ordering wavevector of the
CNC phase are both $\vQ \approx  \vQ_0 -0.14 \pi \vx = 0.86 \pi \vx $.  
The wavevector of the third harmonic $3\vQ $ is then equivalent to $1.42 \pi \vx $ \cite{thr}.

The classical energy $E$ was minimized within a unit cell of length 5,000 with open boundary conditions in the $x$ direction. 
Doubling the unit cell has no noticeable effect on the amplitudes plotted in Figs.2(a) and (b).   In the absence of a
lattice distortion, the amplitudes $B_{2l+1}$ are negligible but the higher harmonics $C_{2l+1 > 1}$ are significant.
For all $D/\vert J_1\vert $, the trial spin configuration has a lower energy than the Monte-Carlo state.  
Notice that the anharmonicity in region 4II is much weaker than in region 4I.   
For the parameters of Fig.2(a), only the third and fifth harmonics $C_3$ and $C_5$ are significant;  for the parameters of Fig.2(b), 
harmonics above $C_7$ can be neglected.  The $S_z$ component of the anharmonic CNC phase within region 4I is sketched 
in the inset to Fig.2(b).  This phase retains some of the Ising character of the $\uudd $ phase with $\langle {S_z}^2\rangle = 0.72$
at $D/\vert J_1 \vert =0.3$..  

Within region 4II, $\vQ=4\pi \vx /3$ depends on neither the exchange parameters nor 
the anisotropy;  within region 4I, $\vQ $ is relatively insensitive to $D$ but sensitively depends on the ratio of exchange parameters,
as discussed in Refs.\cite{Har09} and \cite{Swan09}.   For the parameters used in Fig.2(b), $\vQ \approx 0.857 \pi \vx $.

With decreasing $D$, the anharmonicity decreases in both regions 4I and 4II.  Pure spirals with 
$C_{2l+1 > 1}=0$ and $\langle {S_z}^2\rangle =0.5$ are recovered as $D\rightarrow 0$.  So the phase diagram provided by Fig.3 of 
Ref.\cite{Har09} should be revised to eliminate the sharp boundary between the CNC phase and the spiral region.
Comparing the numerical results obtained for the trial spin configuration
of Eqs.(\ref{sp1}) and (\ref{sp2}) with the earlier Monte-Carlo results \cite{Har09} 
using the parameters of Fig.2(b), we find that the critical value $D_c $ below which the 
$\uudd $ phase disappears increases from $0.295 \vert J_1\vert $ to $0.317\vert J_1\vert $.  When $D/\vert J_1\vert =0.1$, 
the energy of the Monte-Carlo phase is $E/N= -1.284 \vert J_1\vert $ whereas the energy of the anharmonic CNC phase is $-1.295 \vert J_1\vert $.  These energies can be compared with the energy of the $\uudd $ phase in Fig.3(b).

\begin{figure}
\includegraphics *[scale=0.5]{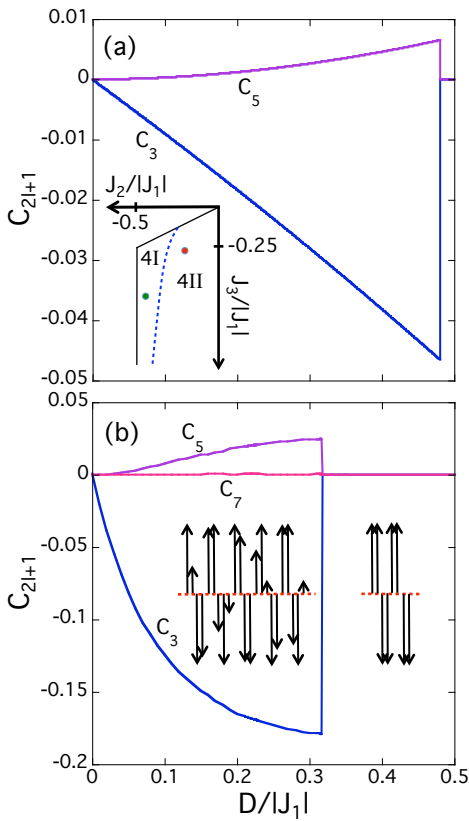}
\caption{
(Color online) The harmonic amplitudes $C_{2l+1}$ as a function of $D/\vert J_1\vert $ for
(a) $J_2/\vert J_1 \vert =- 0.20$ and $J_3/\vert J_1\vert = -0.26$ or 
(b) $J_2/\vert J_1 \vert =- 0.44$ and $J_3/\vert J_1\vert = -0.57$.  Inset in (a) is the phase diagram
indicating regions 4I and 4II discussed in the text with the red dot corresponding to the (a) parameters and
the blue dot to the (b) parameters.  In (b) we sketch the $\uudd $ phase stable above $D/\vert J_1 \vert = 0.32$
and the CNC phase ($S_z$ component only) at $D/\vert J_1\vert =0.3$.
}
\end{figure}

The observed $2\pi \vx -\vQ $ peak in the elastic neutron-scattering measurements \cite{Nak07, Terada08} with 
amplitude $\vert B_1\vert^2$ is absent for a non-distorted lattice.  
This elastic peak requires a lattice distortion with a wavevector of $\vq = 2\pi \vx $ or a period of 1.   
The most likely source of that distortion is the repulsion of neighboring oxygen atoms shown in Fig.1.  
For each pair of oxygen atoms, one lies below the hexagonal layer and the other lies above. The
displacement of oxygen atoms pictured in Fig.1 produces a ``scalene" distortion of the triangular lattice 
that has been observed in both the low-field $\uudd $ phase of pure CuFeO$_2$ as well as in the 
field-induced multiferroic phase above 7 T \cite{Terada06}.  It has also been reported
in the ferroelectric phase of Al-doped CuFeO$_2$ \cite{Nak08}.  The displacement expands the lattice 
in the $x$ direction \cite{expand}, which was observed in pure CuFeO$_2$ \cite{Terada06, Ye06}.

To study the distorted phase, we take $J_1^{(1)}=J_1^{(2)} = J_1 -K_1 /2$
and $J_1^{(3)} = J_1 +K_1$, which ensures that the average bond strength remains $J_1$.  The 
energy $K_1$ measures the degree of distortion of the nearest-neighbor exchange.
When $K_1>0$, the oxygen displacements weaken the AF coupling $J_1^{(3)}$
and strengthen the AF couplings $J_1^{(1)}$ and $J_1^{(2)}$.  The $J_1^{(3)}$ bonds form a zig-zag pattern
with wavevector $2\pi \vx $ or a period of 1.  The lattice distortion lowers the energy of the $\uudd $ phase, which is given by 
$E/N=J_1-J_2+J_3 -D - 2K_1$ for $K_1>0$ and $E/N=J_1-J_2+J_3 -D +K_1$ for $K_1 <0$.   For $K_1> 0$, the 
$\uparrow $ or $\downarrow $ spins prefer to lie on the same zig-zag coupled by the weakest AF bond $J_1^{(3)}$, 
as shown in Fig.1.  By breaking the three-fold degeneracy of the energy, this distortion selects a spin state with 
wavevector along $\vx $ over its two twins with wavevectors rotated by $\pm \pi/3$. 

The amplitudes $B_{2l+1}$ and $C_{2l+1}$ are plotted versus $K_1/\vert J_1\vert $ for $D/\vert J_1\vert =0.1$ in Fig.3(a).
The phase $\phi $ in Eq.(\ref{sp1}) is fixed by the phase of the lattice distortion.
With increasing $K_1/\vert J_1\vert $, the third-order amplitude $C_3$ decreases while the amplitudes $B_1$ and $B_3 $
increase in size.  For $K_1/\vert J_1\vert =0.05$, $B_1\approx -0.21$ and $B_3 \approx -0.05$.  
The anharmonicity of the spin changes with the distortion:  $\langle {S_z}^2 \rangle $ decreases from 0.65 to 0.59 as
$K_1$ increases from 0 to $0.05\vert J_1\vert $.
Spin population profiles $P(S_z)$ for $K_1=0 $ and 
$K_1 = 0.05 \vert J_1 \vert $ are plotted in the inset to Fig.3(a).  
Their difference indicates that
the distorted lattice is more heavily weighted towards $S_z=0$ and
the non-distorted lattice is more heavily weighted towards $S_z = \pm 1$.  As shown in Fig.3(b), the $\uudd $
phase obtains a lower energy than the CNC phase when $K_1 > 0.057 \vert J_1\vert $.

\begin{figure}
\includegraphics *[scale=0.5]{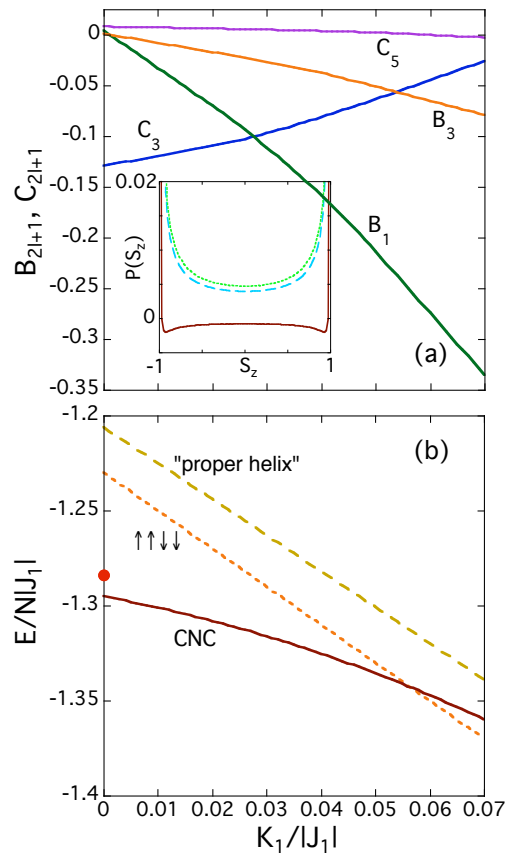}
\caption{
(Color online) (a) The amplitudes $B_{2l+1}$ and $C_{2l+1}$ versus the distortion $K_1$ for 
$D/\vert J_1\vert =0.1$.  Inset are the
spin populations $P(S_z)$ for the spin states at $K_1=0$ (long dash) and 
$K_1/\vert J_1\vert =0.05$ (short dash) as well as their difference (solid).  (b) The energy 
$E/N\vert J_1\vert $ for the predicted CNC phase (solid), the $\uudd $ phase
(short dash) and the ``proper helix" \cite{Nak07} (long dash).    The dot
along the $E$ axis is the energy of the Monte-Carlo simulation \cite{Har09}.
Other parameters as in Fig.2(b).  
}
\end{figure}

Other distortions of the lattice with a period of 1 along $\vx $ can also produce sizeable amplitudes 
$B_1$ at the wavevector $2\pi \vx -\vQ$.  However, the $2\pi \vx -\vQ $ peak cannot be induced by the 
$\vq =0 $ ``isosceles" distortion observed by Feng {\it et al.} \cite{Ye06} in 
pure CuFeO$_2$ with the $J_1^{(1)}$ bonds along the $x $ axis reduced in 
size but the diagonal bonds $J_1^{(2)} = J_1^{(3)}$ remaining identical.

Guided by the observed elastic peaks at $\vQ $ and $2\pi \vx -\vQ $, Nakajima {\it et al.} \cite{Nak07}
constructed a CNC phase different than the one proposed here.  Their ``proper helix" is a modified spiral
with the same spin on sites $\vR = m\vx + n\sqrt{3}\vy $ and $\vR' = \vR + \vx /2 + \sqrt{3} \vy /2$.   
After minimizing its energy as a function of wavevector for $D/\vert J_1\vert =0.1$, we compare
the ``proper helix" with the predicted CNC phase as well as with the pure $\uudd $ 
phase in Fig.3(b).  Not only does
the ``proper helix" have a higher energy than the predicted CNC phase, it also has a higher energy than
the $\uudd $ phase provided that $D/\vert J_1\vert $ is not too small.  
For a non-distorted lattice with $K_1=0$, the ``proper helix" has a 
lower energy than the $\uudd $ phase (but not lower than the predicted CNC phase) when $D/\vert J_1\vert < 0.047$.

Estimating the actual distortion in CuFeO$_2$ requires that we also consider the elastic energy cost
proportional to ${K_1}^2$.  Since the gain in exchange energy in Fig.3(b) is linear in $K_1$, a distortion with period 1
occurs in both the $\uudd $ and the CNC phases.  Allowing a general distortion of the lattice with wavevector 
$\vq $, we find that the energy also has minima at $\vq  = 2\vQ$ and $4\pi \vx - 2\vQ $, 
corresponding to a charge modulation with half the period of the spin modulation.  This modulation has 
been observed \cite{Nak08} in Al-doped CuFeO$_2$ and may be related to the predicted 
ferroelectric instability \cite{Arima07}, which contains both second- and fourth-order harmonics 
in addition to the uniform displacement of the oxygen atoms.  

Of course, it remains possible that the trial spin state used in this work is incomplete and that an even lower-energy
state can be achieved.  But the close agreement with Monte-Carlo simulations \cite{Har09} in Fig.3(b)
bolsters our confidence that the anharmonic CNC state
provides an excellent approximation to the true ground state of Eq.(\ref{Ham}) for classical spins.  As a test of our model,
Fig.3(a) indicates that the multiferroic state should have small elastic peaks at the third harmonics of $\vQ $ 
and $\vQ' $ \cite{thr}.

To summarize, we have shown that the CNC phase of a frustrated triangular lattice contains significant odd-order
harmonics of the fundamental wavevector $\vQ $. This result should greatly facilitate the future modeling of multiferroic
ground states.  As the easy-axis anisotropy $D$ is reduced, the amplitudes of the
higher harmonics are decreased and a pure cycloid is recovered as $D\rightarrow 0$.  The $2\pi \vx -\vQ $
peaks observed in Al- and Ga-doped CuFeO$_2$ are explained by the scalene distortion of the triangular lattice.  

This research was sponsored by the Division of Materials Science
and Engineering of the U.S. Department of Energy.


\begin{references}
\bibitem{thr} In a three-dimensional crystal, the 
instability and ordering wavevectors all have a $z$ component of 3/2 in units of $2\pi /c $,
where the separation between layers is $c/3$.  In $(H,K,L)$ notation,
$\vQ \approx (0.21, 0.21, 3/2)$, $\vQ' = 2\pi \vx - \vQ \approx (0.29, 0.29, 3/2)$, the third harmonic of 
$\vQ $ is equivalent to $(0.37, 0.37, 3/2)$, and the third harmonic of 
$\vQ' $ is equivalent to $(0.13, 0.13, 3/2)$.
\bibitem{expand} An expansion of the lattice in the $x$ direction will also suppress the magnitude
of $J_1^{(1)}$ compared to $J_1^{(2)}$.  However, a uniform change in $J_1^{(1)}$ (with wavevector $\vq =0$)
does not produce any additional elastic peaks.
\end{references}
\end{document}